\documentclass[proceedings]{JHEP} 

\usepackage{epsfig}         

\def\eq#1{(\ref{#1})}
\def\tfrac#1#2{{\textstyle{#1\over #2}}}

\def\be{\begin{equation}}
\def\ee{\end{equation}}
\def\bea{\begin{eqnarray}}
\def\eea{\end{eqnarray}}
\def\barr{\begin{array}}
\def\earr{\end{array}}
\def\bfone{\relax{\rm 1\kern-.35em 1}}

\def\IE{\relax{{\rm I\kern-.18em E}}}
\def\del{\partial}
\newsavebox{\uuunit}
\sbox{\uuunit}
    {\setlength{\unitlength}{0.825em}
     \begin{picture}(0.6,0.7)
        \thinlines
        \put(0,0){\line(1,0){0.5}}
        \put(0.15,0){\line(0,1){0.7}}
        \put(0.35,0){\line(0,1){0.8}}
       \multiput(0.3,0.8)(-0.04,-0.02){12}{\rule{0.5pt}{0.5pt}}
     \end {picture}}
\newcommand {\unity}{\mathord{\!\usebox{\uuunit}}}

\newbox\mybox
\newcommand\fverb{\setbox\mybox=\hbox\bgroup\verb}
\newcommand\fverbdo{\egroup\medskip\noindent\fbox{\unhbox\mybox}\ }
\newcommand\fverbit{\egroup\item[\fbox{\unhbox\mybox}]}

\font\beeg=cmr17 scaled 1600        
\newcommand\init[1]{\setbox\mybox=\hbox{{\beeg #1}~}%
           \noindent\global\hangindent=\wd\mybox\global\hangafter-2%
           \sc\smash{\llap {\lower 13.2pt \box\mybox}}}

\title{Embedding Branes in Flat Two-time Spaces}

\author{L. Andrianopoli$^a$, M. Derix$^a$,
 G.W. Gibbons$^b$,
  C. Herdeiro$^b$,
 A. Santambrogio$^a$,
A. Van Proeyen$^{a}$ \thanks{Onderzoeksdirecteur, FWO, Belgium}\\
$^a$ Instituut voor Theoretische Fysica, Katholieke
 Universiteit Leuven,\\
Celestijnenlaan 200D B-3001 Leuven, Belgium
\\ \vspace{6pt}
$^b$ D.A.M.T.P., University of Cambridge,\\ Silver Street, Cambridge CB3
9EW, U.K.}

\conference{Quantum aspects of gauge theories, supersymmetry and
unification, Paris 1999}

\abstract{We show how non-near horizon, non-dilatonic $p$-brane theories
can be obtained from
two embedding constraints in a flat higher dimensional space with 2
time directions. In particular this includes the construction of D3
branes from a flat 12-dimensional action, and M2 and M5 branes from
13 dimensions. The worldvolume actions are found in terms of fields
defined in the embedding space, with the constraints enforced
by Lagrange multipliers.
}

\begin{document}

\maketitle 

{\init Embedding} geometric manifolds in higher dimensional flat
spaces can be a useful tool for studying global properties of
these manifolds. Typical examples of extending the dimension for a
better understanding of the geometry are the description of the
$d$-dimensional sphere $S^d$ by embedding it in $(d+1)$-dimensional
Euclidean space and the description of $AdS_d$ as a hyperboloid in
flat $(d+1)$-dimensional space with two timelike directions. The
embedding is encoded, in both cases, in one constraint on the
embedding coordinates and makes the $SO(p,q)$ global symmetry \footnote{$SO(d+1)$
for the sphere, $SO(d-1,2)$ for AdS} of these
geometries manifest. Any choice of coordinates on the
$d$-dimensional manifold will break this manifest symmetry.

The two examples above are combined in \cite{conffads}, where the
$AdS_{p+2}\times S^{n-1}$ near-horizon geometry of $p$ branes in
$D=p+n+1$ dimensions is described starting from a flat
$(D+2)$-dimensional space. Two constraints are imposed, which
respectively reduce $p+3$ dimensions to the $AdS_{p+2}$ manifold and
$n$ dimensions to the $S^{n-1}$ sphere. The Born--Infeld action for
the near-horizon theories of these branes can then be expressed in
terms of
$(D+2)$-dimensional fields, while the embedding constraints are imposed
by means of Lagrange multipliers. The Wess--Zumino
(WZ) terms in these actions can be obtained from a constant $(p+2)$
form in $D+2$ dimensions, which must be
integrated over a $(p+2)$-dimensional manifold which has the
worldvolume as its boundary.

This talk is based on the paper \cite{noi}. We will show,
following \cite{noi}, how the full spacetime metric of a brane can
be embedded isometrically in ${\IE}^{D,2}$. In this way we
generalize the construction of \cite{conffads} to not just
near-horizon, but to the full brane geometry. We will show that even
if the geometry is not a product of $AdS$ times a sphere, the
brane geometry can still be embedded in flat $(D+2)$-dimensional
space with signature $(D,2)$. The two constraints are no longer
independent in the sense that they do not constrain separate
coordinates of the embedding space, but instead a non-trivial
mixing of the coordinates is involved.

As in \cite{conffads}, also the forms for the WZ terms are
obtained in this picture. For that construction, we will assume
\cite{hew} that a $p$ brane evolving in a space with two times
couples to a $(p+3)$-form field strength. The field strength is
then contracted to a $(p+2)$-form which can be used for the WZ
term. To make this contraction we will have to introduce an extra
vector field which only in the near-horizon limit will have an
elegant form.

One may wonder whether the whole geometry cannot be embedded with
just one extra dimension and why we need  two timelike directions
for the embedding space. First of all, it has been shown
\cite{eise} that the embedding of a surface in a flat space of
co-dimension 1 imposes, by use of the Einstein equations, that the
surface has constant curvature (if the surface has dimensionality
$d>2$). This corresponds to familiar cases as the embedding of
spheres and (anti--)de Sitter in flat spaces. Therefore, in order
to embed brane backgrounds, which do not have constant curvature,
we need at least two extra dimensions.

To determine the signature of the embedding space we use the
following argument. An interesting aspect of brane spacetimes is
that they are not globally hyperbolic\footnote{A space is called
globally hyperbolic if it possesses a Cauchy surface}. According
to Penrose~\cite{Penrose}, a global isometric embedding in normal flat
Minkowski space, i.e. in ${\IE}^{n,1}$, is only possible for a
spacetime which is globally hyperbolic. Penrose's argument is
essentially that the restriction of the time coordinate $X^0$ of
${\IE}^{n,1}$ to the embedded spacetime $M$ would serve as a
time-function on $M$, i.e., a function which increases along every
future directed timelike curve. Moreover, if the embedding is
suitably regular, the level sets (constant time slices) would
actually serve as Cauchy surfaces on $M$, implying global
hyperbolicity. In a sense any embedded surface inherits global
hyperbolicity from the ambient space. No such obstruction arises
for embeddings in flat space with more than one timelike direction.
We will describe therefore a minimal embedding of general brane
backgrounds in flat spaces with two extra dimensions and $(D,2)$
signature.

In section~\ref{ss:embedding} we give the embedding of the
geometry, commenting on global properties and on the near-horizon
approximation. The worldvolume actions will be constructed in
section~\ref{ss:braneaction}. The essential step in that section
is the construction of the forms. General results for the electric
field strengths are given, before completing the construction for
the example of the  D3-brane.


\section{Embedding: The geometry}
\label{ss:embedding} In this section we describe the embedding of
a $SO(n)$ invariant $p$-brane in a $(D+2)=(n+p+3)$-dimensional
spacetime. We will obtain the embedding by requiring that the
known metric of the brane is obtained from a flat $(D,2)$ metric,
i.e. we demand that the embedding is isometric
\cite{goenner,friedman}. The $D$-dimensional $p$-brane geometry
can generally be described by a metric of the form
\bea
ds^{2}&=&A(r)^{2}\left[ -dt^{2}+dx_{p} \cdot
dx_{p}\right]+\nonumber\\
&&+B(r)^{2}dr^{2}+
C(r)^{2}d\Omega_{n-1}^{2}\,,
\label{metricbra}
\eea
where $dx_p\cdot dx_p$ is the $p$-dimensional spacelike part of the
worldvolume, and $d\Omega_{n-1}^{2}$ is the metric for the $n$-sphere.

On the embedding space $\IE^{(D,2)}$ we take cartesian coordinates
$X^{M}$, with
 $M=0,\ldots ,D+1$, which we divide as
$X^M=(X^\mu, X^{p+1},  X^{p+2}, X^\alpha)$ (with $\mu =0,\ldots
,p$, $\alpha=p+3,\ldots ,D+1$). Using these coordinates, the metric
looks as follows
\bea
ds^{2}&=&-(dX^{0})^{2}+(dX^{1})^{2}+\ldots+(dX^{p+1})^{2}+
\nonumber\\
&&-(dX^{p+2})^{2} +\ldots+(dX^{D+1})^{2}\,.
\eea

To obtain the two embedding constraints, we start by making a change of
the $(D+2)$-dimensional coordinates,
such that we make a subgroup $SO(p,1)\times SO(n)\subset
SO(p+n+1,2)$ manifest. This is achieved by using a mixture of hyperspherical
 and horospherical coordinates $\{\rho ,z,x^\mu ,\beta ,n^\alpha \}$
\begin{eqnarray}
&& X^-\equiv X^{p+2}-X^{p+1}=\frac{\rho}{z},\nonumber\\
&&X^+ \equiv X^{p+2}+X^{p+1}= \rho z+\frac{\rho}{z}x^{\mu}x_{\mu},\nonumber\\
&& X^{\mu}=\rho \frac{x^{\mu}}{z}\,,\quad X^{\alpha}=\beta n^{\alpha}\,,
\label{embori}
\end{eqnarray}
where $n^{\alpha}$ ($\sum_{\alpha}(n^{\alpha})^{2}=1$) parametrise
 the sphere $S^{n-1}$.
In these new coordinates, the metric reads
\be
ds^{2}=\frac{\rho^{2}}{z^{2}}\left[dx^\mu dx_\mu +dz^{2}\right]-d\rho^{2}
+d\beta^{2}+\beta^{2}dn ^\alpha dn^\alpha \,.
\label{metricemb}
\ee

Comparing (\ref{metricemb}) and (\ref{metricbra}) we
identify $dx^\mu dx_\mu$ with $-dt^2+dx_p.dx_p$ and $dn^\alpha dn^\alpha
$ with $d\Omega _{n-1}^2$. Then $\beta $, $\rho $ and $z$ are
functions of $r$ and are still to be determined. The comparison gives
\begin{eqnarray}
&& \beta=C(r) \, ,\quad \frac{\rho}{z}=A(r) \, ,\nonumber\\
&&
-d\rho^{2}+\frac{\rho^{2}}{z^{2}}dz^{2}+d\beta^{2}=B(r)^{2}dr^{2}\,.
\label{ident}
\end{eqnarray}
The differential equation can be rewritten to give
\be
\frac{C'^{2}-B^{2}}{A'}=(\rho z)' \equiv F'\,.
\label{df}
\ee

{}From all this we can derive the following embedding functions
\begin{eqnarray}
X^- &=& A(r) \, , \quad
X^+ = F(r)+A(r)x^{\mu}x_{\mu}\nonumber\\
X^{\mu}&=&A(r)x^{\mu} \, , \quad
X^{\alpha}=C(r) n^{\alpha}\,.
\label{embedfun}
\end{eqnarray}
We can, furthermore, express the constraints in terms of the $X^M$
coordinates only. Denoting the inverse function with an overbar,
i.e., $\bar{f}(f)=f(\bar{f})=\unity $, we can write
$r=\bar{A}(X^-)$. Thus, our two constraints are
\begin{eqnarray}
\phi_1 &=& X^-X^+-X^{\mu}X_{\mu}-
X^- F(\bar{A}(X^-))=0\nonumber\\
\phi_2 &=& \sum_{\alpha}(X^{\alpha})^{2}-
\left[ C(\bar{A}(X^-))\right]^{2}=0\,.
\label{constraints}
\end{eqnarray}
These constraints are therefore determined by the functions $A$, $C$ and
$F$. The latter is determined up to a constant by (\ref{df}) in terms
of $A$, $B$ and $C$.

Note that so far there is no definition of the
radial variable $r$. We can use different parametrizations,
e.g. it will turn out that in some cases it is useful to take $A$ or $C$
itself as the radial variable.
In the standard brane cases, the functions $A$, $B$ and $C$ will take
the form of some harmonic function to some power
in the transverse space of the brane. We will further adopt the name
$r$ for that transverse coordinate, use just the name $A$ for the
parameter in the first mentioned parametrization, and use $R$ for the
radial coordinate such that $C(R)=R$.

{}From now on we will assume that the functions $A$, $B$ and $C$ are
indeed harmonic functions in $n$ dimensions with a flat limit at
$r\rightarrow \infty$.
For non-dilatonic D- and M-branes, they are of the
following form
\begin{eqnarray}
&H =   \left(1+\frac{1}{r^{\kappa}}\right)\,,\quad
A(r)  =  H^{-\frac{1}{p+1}}&\nonumber\\
&B(r)  =  H^{\frac{1}{\kappa}}\,,\quad
C(r)= r H^{\frac{1}{\kappa}}&
\label{harmonic}
\end{eqnarray}
where $\kappa \equiv n-2=D-p-3$.
Here a priori $r>0$ and $r=0$ corresponds to the horizon, but we will
come back to this later.

With this explicit form for the functions $A$, $B$ and $C$ we can evaluate
the function $F$. Using (\ref{df}) we get
\be
F'(r) = - w r^{1-\kappa}
(1+r^{-\kappa})^{\frac{2}{\kappa} + \frac{1}{p+1} -1} (1+2r^\kappa)
\,,
\ee
(where $w=\frac{p+1}{\kappa}$), which can be integrated to give
(up to a constant)
\bea
F(r) &=& - \tfrac{w}{\kappa} \left[ B_\frac{r^\kappa}{r^\kappa
+1}\left(\tfrac{-1}{p+1}, 1-\tfrac{2}{\kappa}\right) \right. + \nonumber\\
&& \left. + 2 B_\frac{r^\kappa}{r^\kappa+1} \left( \tfrac{p}{p+1},
 -\tfrac{2}{\kappa}\right) \right] \,.
\label{fofr}
\eea
Here we used the incomplete Beta function
\bea
 B_x(a,b) &=& \int_0^x  t^{a-1} (1-t)^{b-1} dt=\nonumber\\
&=& a^{-1}x^a {}_2F_1(a,1-b;a+1;x)\,, \nonumber
\eea
which is defined for $0 < x \leq 1$. This means that $F(r)$ is well
defined in the region $r>0$, which is what we were looking for.

Note that near the horizon ($r\to 0$), where the brane geometry is well
described by $AdS_{p+2}\times S_{D-p-2}$ \cite{interp}, we get
$ F\sim w^2 r^{-\frac{1}{w}}$
 and the embedding functions (\ref{embedfun}) reduce to those used in
\cite{conffads}.

Using the embedding \eq{embedfun}
 we can now study the global properties of the
brane geometries. Before considering the higher dimensional D- and
M-branes, let us first look at the
simpler example of the extreme Reissner--Nordstr{\o}m (RN) black hole. (A large
list of embedding functions for other solutions of General Relativity
is given in \cite{rosen}).
The RN black hole fits our general embedding scheme with $D=4$ and
$p=0$, $\kappa =1$, $w=1$. Here, rather than working with the radial
variable $r$ as in (\ref{harmonic}), we use the variable $R\equiv r+1$,
which has the
property $C(R)=R$. Then the functions $A$ and $B$ are given by
$A(R) = B(R)^{-1} = 1-1/R$. The horizon is now at
$R=1$ and $R=0$ corresponds to the singularity.
Using (\ref{df}), we then find
\[ F_{RN}(R) = \frac{1}{R-1} -3R - R^2 - 4 \log |R-1| \,.\]

The entire Reissner--Nordstr{\o}m black hole geometry can be drawn
using parametrization (\ref{embedfun}) as is shown in
figure \ref{rn}. We only draw the relevant directions ($X^-$,
$X^+$, and $X^0$), which basically means we only draw the $R$ and
$t$ coordinates of the black hole (every point in the graph should
be thought of as a 2-sphere). The lines in the graph are therefore
constant $t$ and constant $R$ lines.

We can read off the following global features from the picture. The geometry
consists of 2 distinct regions: region I, the asymptotically flat region for
$R>1$ which corresponds to $X^->0$. For big $R$ the surface flattens
and $X^- \to 1$, which is the flat limit. Region
II is the region inside the horizon ($X^- <0$), the singularity ($R=0$)
corresponds to $X^- \to -\infty$. This is the global picture we recognize
from the familiar Penrose diagram for extreme RN black hole as can be
found in \cite{noi}.

The two regions are  connected in an {\it $AdS$-throat}.
It seems that these two
regions are disconnected, the constant time lines all diverge near
$X^-=0$ and never cross the horizon, but this is just
an artifact of the parametrization. Actually, we know that the
near-horizon geometry is equivalent to $AdS_2$, which is known
to have no problems at its 'horizon'. Indeed, a different parametrization
exists (the advanced or retarded Finkelstein coordinates) in which
lightlike geodesics pass smoothly through the horizon into
the interior region, as is depicted in
figure \ref{rnv}.

\EPSFIGURE{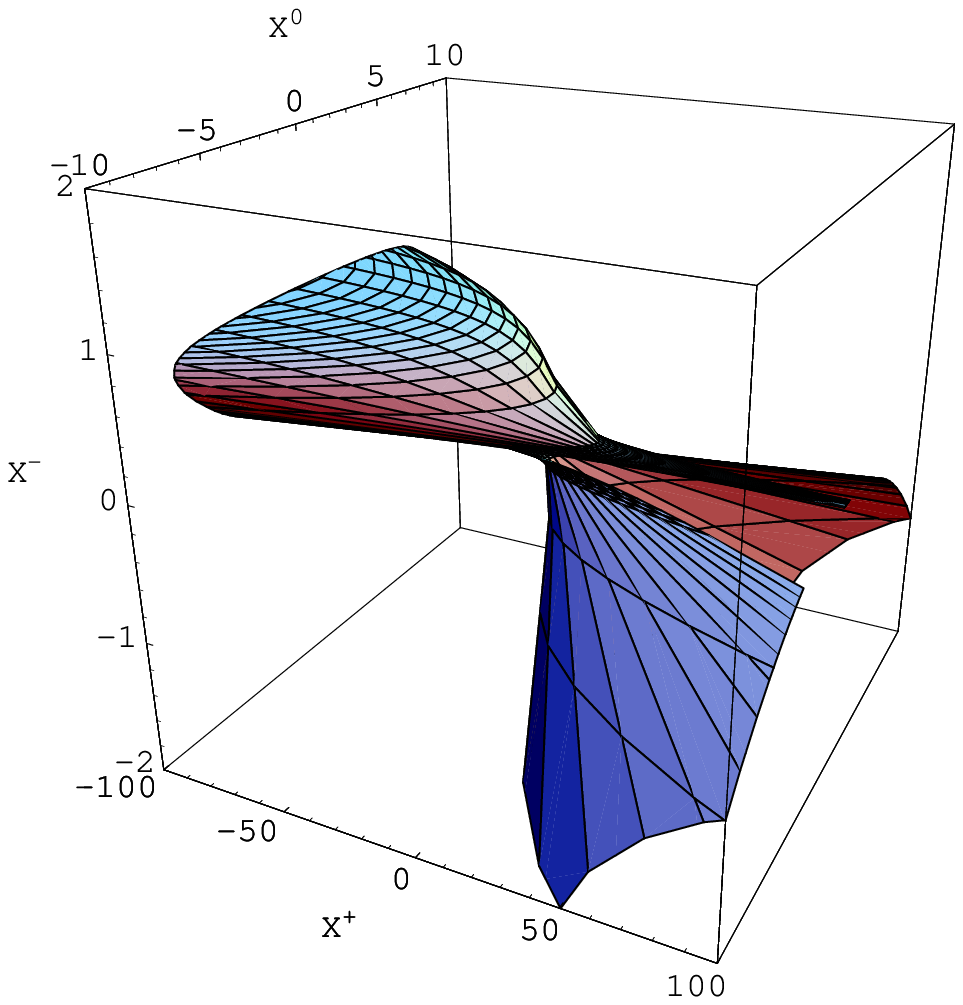, width=20em}{\it Extreme Reissner--Nordstr{\o}m black hole
  parametrised by $R$ and $t$.\label{rn}}
\EPSFIGURE{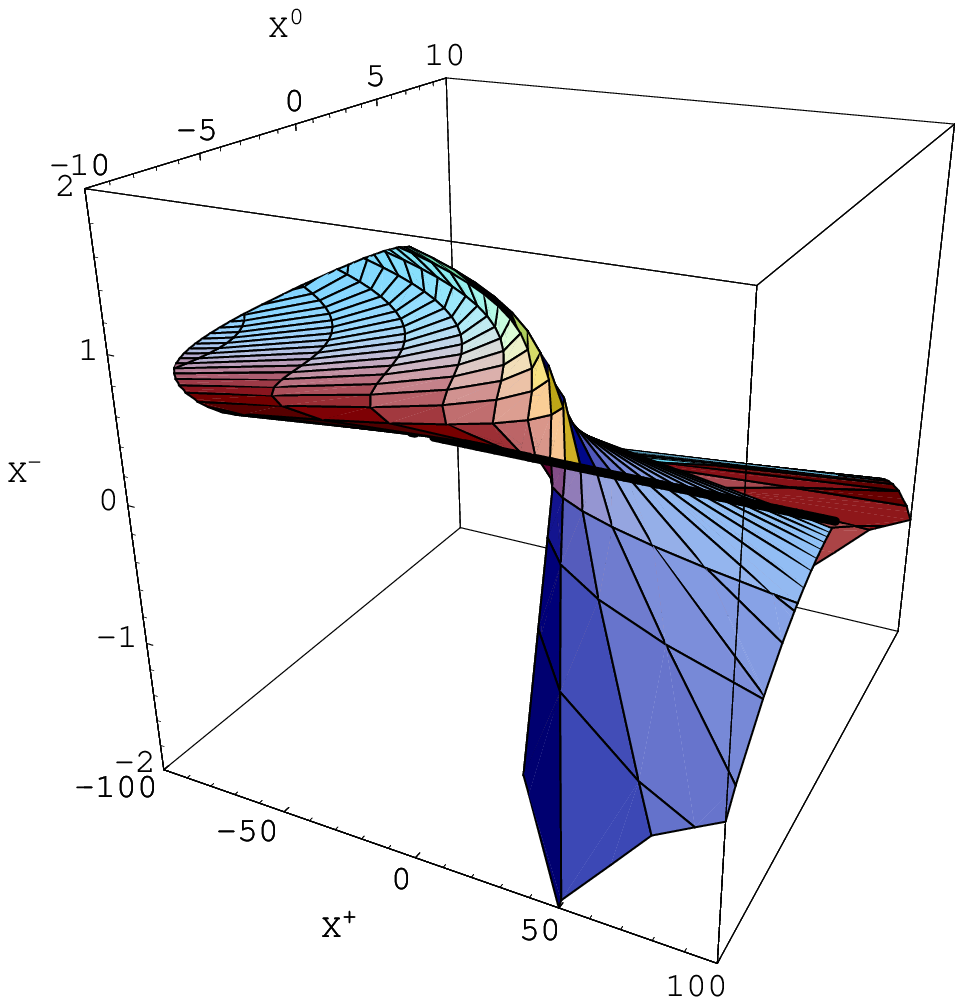, width=20em}{\it Extreme Reissner--Nordstr{\o}m black hole
 in advanced Finkelstein coordinates $R$ and
$v$ (with $v=t-R^*$, where $R^*=\int \frac{B(R)}{A(R)}dR$).
The thick horizontal line is the horizon $R=1$.\label{rnv}}

One of the features of $AdS$ spaces is that they admit closed timelike
curves. The usual
remedy for this is to consider the covering space $CAdS$ instead
of $AdS$ itself. Looking at figures \ref{rn} and \ref{rnv} we see that the RN
black hole geometry suffers from the same problem, it admits closed timelike
curves. Again this is remedied by considering the covering
space. The result of this of course is that the space then consists of
multiple universes.

Let us now move to the non-dilatonic branes.
As discussed in \cite{GHT}, the general brane solution case
(\ref{harmonic}) can be divided in two classes: $p$ odd or $p$ even,
with quite different global properties.

Let us first consider the $p$ \emph{odd} case. In the exterior region ($r>0$),
the function $A(r)$ is analytic and positive and vanishes as
$r\to 0$. If we take $A$ to be our new radial variable instead of $r$,
we see that $A$ can be continued through the horizon to negative $A$
\cite{GHT}. The range of $A$ is from -1 to~1.
The analytic extension of the metric is
\bea
&&  ds^2= A^2 dx_\mu dx^\mu +
\left(1- A^{p+1}\right) ^{-\frac{2}{\kappa }}\times \nonumber\\
&\times &  \left[ w^2\left(1- A^{p+1}\right) ^{-2}A^{-2}dA^2+d\Omega ^2\right]
 \,,
\label{dsA}
\eea
which is even in $A$. This leads to
\bea
F(A)&=& -  \tfrac{w}{\kappa}  (\mbox{sign }A) \bigl[ B_{A^{p+1}}
\left(\tfrac{-1}{p+1}, 1-\tfrac{2}{\kappa}\right) + \nonumber\\
&& + B_{A^{p+1}} \left( \tfrac{p}{p+1}, -\tfrac{2}{\kappa}\right)
\bigr]\,.
\eea
The embedding functions (\ref{embedfun}) are then odd in $A$.
This means that
the embedded space is symmetric around the horizon and completely
nonsingular.
For the non-dilatonic branes, the D3 and M5 fit this picture.
The embedding, depicted in figure \ref{d3m5} for the D3-brane case,
nicely shows these features (an analogous picture for the M5-brane can be found
in \cite{noi}).
It is clearly visible there is no interior region, just two symmetric
'exterior' regions connected in the AdS-throat as was expected from
the Penrose diagram \cite{GHT} \cite{noi}.

\EPSFIGURE{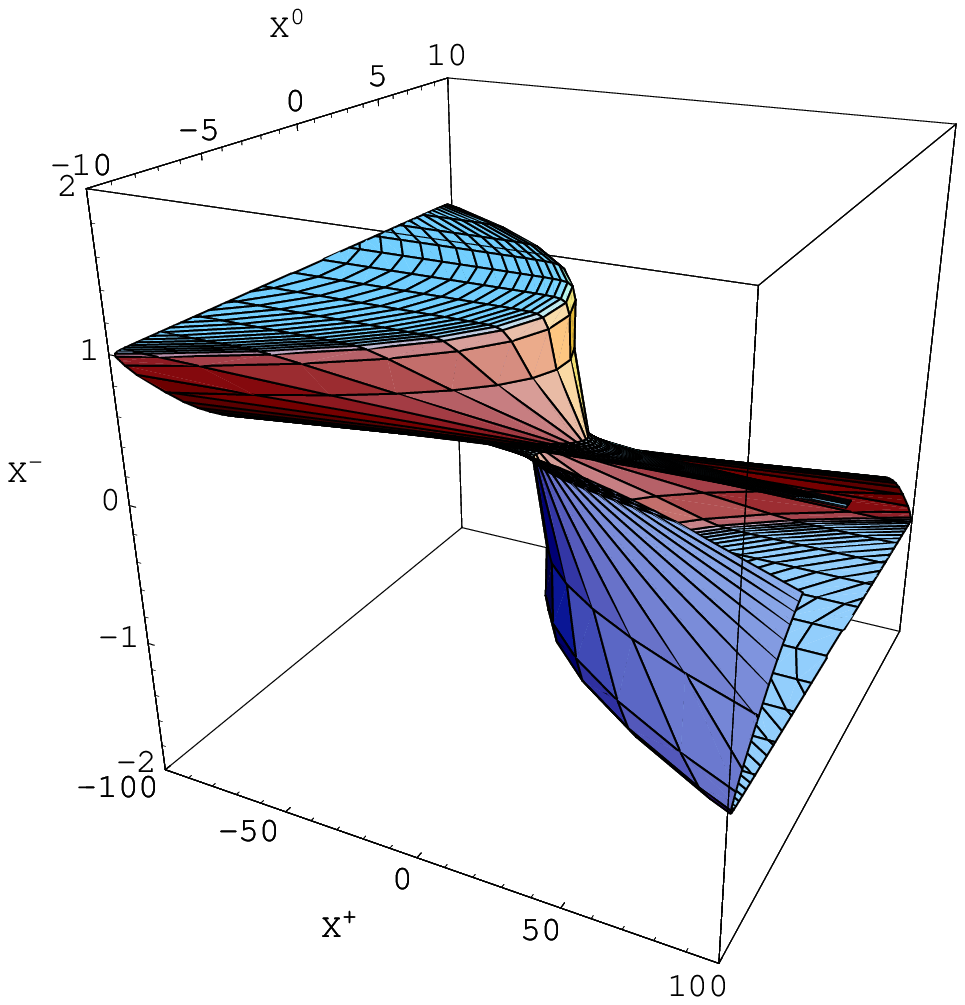,width=20em}{\it $p$ odd branes (D3).
 Two asymptotically flat regions connected in AdS. \label{d3m5}}

In the $p$ \emph{even} case, the metric and embedding functions are neither
even nor odd. It is useful in this case to adopt so-called
Schwarzschild coordinates, defined by $R^\kappa =r^\kappa+1$.
In these coordinates the horizon (which is still a coordinate
singularity) is at $R=1$. At $R=0$ there is a true curvature
singularity.
Expressed in this coordinate, $A(R)$ can be continued
through the horizon into negative $A$ and its range is
$\{-\infty,1\}$. As already stated in \cite{GHT}, the Penrose diagram
for these spaces is equivalent to the extreme Reissner--Nordstr{\o}m diagram.

The embedding of the M2-brane metric illustrates these features.
The expression (\ref{fofr}) of $F$ is only well
defined in the region $R>1$. It is not possible to find a
continuous
expression for $F$ valid in both regions ($0<R\leq 1$ and
$R>1$). But, nevertheless, a continuous embedding is obtained using in the interior region
\bea
 F(R<1) &=& \tfrac{w}{\kappa} \bigr[ B_{R^\kappa}\left(
\tfrac{1}{p+1} + \tfrac{2}{\kappa}, -\tfrac{1}{p+1}\right)+\nonumber\\
&& - 2 B_{R^\kappa}\left( \tfrac{1}{p+1} +
\tfrac{3}{\kappa},-\tfrac{1}{p+1}\right) \bigr]\,. \nonumber
\eea
The global properties of the M2-brane are qualitatively the same as those of the RN
black hole depicted in figure \ref{rn}. We refer to \cite{noi} for the M2-picture.

\section{The brane action}\label{ss:braneaction}
We would like to write the action of a brane placed in the background of
other branes using the embedding of the previous section.
A typical (schematic) form of the action is
\begin{eqnarray}
S_{p+1}&=& \int_W d^{p+1}\xi \sqrt{-\det {\cal G}_{\mu \nu }} +
\int_{B} \Omega _{(p+2)}  +\nonumber\\
&& + \int_W d^{p+1}\xi [\lambda_1 \phi_1 +
\lambda_2 \phi_2]\,,
\label{action}
\end{eqnarray}
where $W=\partial B$ is the $(p+1)$-dimensional world volume of the brane.
The expression for ${\cal G}_{\mu \nu }$ differs in each case.
For example, for Dp-branes ${\cal G}_{\mu\nu} \equiv  \partial _\mu X^M \partial _\nu
 X^N \eta_{MN}+{\cal F}_{\mu\nu}$, with
 ${\cal F}_{\mu\nu}$ the field strength of the gauge field living on the world volume
of the brane. The fields  $\lambda_1,
\lambda_2$ are two Lagrange multipliers implementing the constraints
(\ref{constraints}). $\Omega _{(p+2)}(X^M)$ is a function of
the forms coupling to the brane, such that it reduces to the
appropriate Wess--Zumino term when projected onto the physical
hypersurface.
Its explicit form will be determined for the
D3-brane case in section \ref{ss:d3fs}.
An analogous treatment for M2 and M5 is given in \cite{noi}.

\subsection{Embedding the field strength}
Let us now try to embed the
field strengths appearing in the Wess--Zumino term.
We will assume \cite{hew} that a brane (extended in $p$ spatial directions)
fluctuating in a spacetime with two times
should evolve in both time directions, and therefore couple to
a (p+3)-form field strength.
We assume therefore that the $(D+2)$-dimensional theory can be coupled to
a rank $p+3$ electric field strength $K_e$, and to a rank $n$ magnetic
field strength $K_m$.

This ansatz is the most natural one for the D3-brane,
because in this case the $10$-dimensional self-dual field strength is extended
to a self-dual field strength in $12$ dimensions.
If there would be a supergravity theory in
$D=12$, the bosonic configuration with flat
$(10,2)$ space and a constant self-dual  field strength
would solve the equations of motion. This is obvious for
the Maxwell equation (there can be no Chern--Simons terms built from a 5
form potential in 12 dimensions and so the Maxwell equation would take
the standard form), but for the Einstein equations it is only true because
the field strength is self-dual. In a $D$-dimensional spacetime with
 zero or two times, a self-dual field strength has a vanishing energy momentum
tensor for $D= 4$ mod 4.
(For Lorentzian signature it is $D=2$ mod 4).
What this would mean is that the ten-dimensional D3-brane solution would
just be the projection to a complicated hypersurface of an almost trivial
12 dimensional supergravity solution.

Let us start by analysing how an electric $(p+2)$-form
 field strength $F^{(p+2)}$ gets
embedded in the $(D+2)$-dimensional space.
Our aim is to obtain $F$ as a restriction of a $p+3$-form $K^{(p+3)}$ to the
$D$-dimensional hypersurface $\Sigma$.
A general non-dilatonic brane is described in $D$ dimensions by the fields
 \cite{d3brane}
\begin{eqnarray}
ds^{2}&=&H^{-\frac{2}{p+1}}\left[ -dt^{2}+dx_1^{2}+ \cdots +dx_p^{2} \right]
 +\nonumber\\
&&+H^{\frac{2}{\kappa}}\left[ dr^{2} +r^2 d\Omega_{D-p-2}^{2}\right],
\nonumber\\
G_{01...p}&=&-H^{-1}=-A^{p+1},
\nonumber\\
\Phi&=&0\,,
\label{fields}
\end{eqnarray}
using the notation of section~\ref{ss:embedding}.
We can write the (electric) field strength as
($F\equiv dG$)
\be
F = -(p+1) A^p A' dr\wedge dt\wedge dx_1\wedge \cdots \wedge dx_p\,,
\label{fieldstrength}
\ee
where the prime denotes differentiation with respect to $r$.

To find the embedding,
we start by considering a constant  $(p+3)$-form in $D+2$ dimensions
\begin{equation}
K_e=\frac{p+1}{(p+3)!}\epsilon_{\mu'_{0}...\mu'_{p+2}}
dX^{\mu'_{0}}\wedge dX^{\mu'_{1}}...\wedge dX^{\mu'_{p+2}}
\label{p+3form}
\end{equation}
($\mu'=0,\ldots,p+2$). In order to get a rank $(p+2)$ field strength, we contract $K_e$ with a vector
 field $V$, with components $V=V^M(\frac{\partial}{\partial X^M}$),
 which so far remains arbitrary. (There is a sign ambiguity in this contraction;
 we chose to make it on the left, i.e. $K(V)_{\mu'_{1}...\mu'_{p+2}}\equiv
V^{\mu'_{0}}K_{\mu'_{0}...\mu'_{p+2}}$). Such a contraction yields
\begin{equation}
K_e(V)=\frac{p+1}{(p+2)!}\epsilon_{\mu'_{0}...\mu'_{p+2}}V^{\mu'_{0}}
dX^{\mu'_{1}}\wedge ... dX^{\mu'_{p+2}}\,.
\label{p+2form}
\end{equation}
Then we reduce the resulting $(p+2)$-form to the $D$ dimensional hypersurface
 by using the embedding functions (\ref{embedfun}),
\begin{eqnarray}
&K_e(V)|_{\Sigma} =
 \frac{p+1}{2}A'A^{p+1}  dr\wedge dt \wedge dx_1 \wedge \cdots  dx_p \times
 \nonumber\\
&\times
 \left[ 2V^{\mu}x_{\mu}+ V^{+}(\frac{F'}{A'}-x^{\mu}x_{\mu})
-V^{-}\right]\,,
\label{hev}
\end{eqnarray}
where we defined $V^{\pm} \equiv V^{p+2}\pm V^{p+1}$.
Next we impose that $K_e(V)|_\Sigma = F$.
{}From this we can determine $V^{M}$, using the ansatz
 $V^{\mu'}=\alpha(r)X^{\mu'}$. Because
$K_e$ only has components in the longitudinal directions, $V^\alpha$
stays undetermined. When the field strength also includes a magnetic
part, this $V^\alpha$ comes into play, as we will see in the next subsection.
It follows that, in order for (\ref{hev}) to match with (\ref{fieldstrength}),
$\alpha (r)$ has to obey
\be
\alpha(r)(\frac{AF'}{A'}-F)=-\frac{2}{A}
\label{alpha}
\ee
which gives, using (\ref{df})
\be
\alpha(r)=\frac{2}{AF+w^2 C^2 (2C^{\kappa}-1)}\,.
\ee
or, in terms of the embedding coordinates ($\alpha(r)\to \alpha(r(X))\equiv
\alpha(X)$)
\be
\alpha(X)= \frac{2}{w^2 (X_{\alpha})^2
\left[2(X_{\alpha})^{\kappa}-1 + w^{-2}
\right] -(X_{M})^2}
\ee
and $V^{\mu'}(X)=\alpha (X)X^{\mu'}$.

\subsection{D3-brane embedding}
\label{ss:d3fs}
Let us now discuss, as an example, how this construction works for the
D3-brane in the background produced by other D3-branes.
We refer to \cite{noi} for a discussion of the embedding of the M2-
and M5-brane.

The 10-dimensional Wess--Zumino term is the integral of the self-dual field strength $F$
that couples to the D3-branes solution of the type IIB supergravity theory.
For the 12-dimensional theory we construct a self-dual 6 form $K$, i.e.
\begin{equation}
\star{K}\wedge K = \eta_{12}|K|^{2},
\end{equation}
where $\eta_{12}$ is the volume form on $\IE^{(D,2)}$.
Our aim is to obtain $F$ as a restriction of $K$ to the 10 dimensional surface $\Sigma$.
The D3-brane is described by the fields (\ref{fields}) with $p+1=\kappa=4$, $D=10$.
We can therefore write the self-dual field strength
in terms of the embedding functions as
\begin{equation}
F =- 4 A'A^{3}dt\wedge dx\wedge dy \wedge dz \wedge dr
 + 4 \omega_{(5)} \, ,
\label{fieldstr}
\end{equation}
where $\omega_{(5)} \equiv
\sin(\theta)^{4}\sin(\phi_{1})^{3}\sin(\phi_{2})^{2}\sin(\phi_{3}) d\theta \wedge d\phi_1 \wedge ... \wedge d\phi_4$ is the volume form on the
unit 5-sphere.

To find the embedding, we again start by considering a constant form in the embedding space,
which in this case we take to be a self-dual six-form
\bea
K&=&\frac{4}{6!}(\epsilon_{\mu'_{0}...\mu'_{5}}dX^{\mu'_{0}}\wedge dX^{\mu'_{1}}...\wedge dX^{\mu'_{5}} +\nonumber\\
&+& \epsilon_{\alpha_{1}...\alpha_{6}}dX^{\alpha_{1}} \wedge dX^{\alpha_{1}}...\wedge dX^{\alpha_{6}}),
\label{sixform}
\eea
In order to get a rank 5 field strength, we contract, as we have done for the
general electric case, $K$ with a vector field $V$.
Such a contraction yields
\bea
K(V)&=&\frac{4}{5!}(\epsilon_{\mu'_{0}...\mu'_{5}}V^{\mu'_{0}} dX^{\mu'_{1}}\wedge ...\wedge dX^{\mu'_{5}} +\nonumber\\
&+& \epsilon_{\alpha_{1}...\alpha_{6}}
V^{\alpha_{1}} dX^{\alpha_{2}}\wedge ...\wedge dX^{\alpha_{6}})\,.
\label{fiveform}
\eea
Again we reduce the resulting 5 form to the 10-dimensional hypersurface by using the embedding functions
(\ref{embedfun}). By requiring the matching $K(V)|_{\Sigma}=F$, we get the constraints on our
vector field $V$. The resulting 5 form $K(V)|_\Sigma$ is precisely the Wess--Zumino
term $\Omega_5$ we were looking for.

Let us analyse separately the two terms in the right hand side of
(\ref{fieldstr}), (\ref{sixform}) and (\ref{fiveform}).
The electric part has already been studied in the general case in the
previous subsection. In this case it gives $V^{\mu'}=\alpha(r) X^{\mu'}$
with
\begin{equation}
  \alpha (X)=\frac{2}{2 (X^{\alpha}X_{\alpha})^3  -X^{M}X_{M}}\,.
\end{equation}
The magnetic part in (\ref{sixform}) can be rewritten in terms of the radial
coordinate $r$ and the angular coordinates $\theta$,$\phi_i$ ($i=1,\cdots
,4$)
\begin{eqnarray}
\frac{1}{6!} \epsilon_{\alpha_{1}...\alpha_{6}}dX^{\alpha_{1}} \wedge \cdots
\wedge dX^{\alpha_{6}} =\nonumber\\
\quad\quad = C' C^5 dr \wedge \omega_5\,.
\end{eqnarray}
For the second term in (\ref{fiveform}) to match with the second
 term in (\ref{fieldstr}), we have to require that the vector $V^\alpha$ points in the
 radial direction when decomposed in the $r,\theta , \phi_i$ basis, that is
\be
V^\alpha \frac{\del }{\del X^\alpha} \equiv V^\alpha \frac{\del r }{\del X^\alpha}
\frac{\del }{\del r}\,.
\label{radialv}
\ee
This gives
\bea
\tfrac{1}{5!} \epsilon_{\alpha_{1}...\alpha_{6}}V^{\alpha_{1}}  dX^{\alpha_{2}}
\wedge ... \wedge dX^{\alpha_{6}} = \nonumber\\
\quad\quad = C' C^5 V^\alpha \frac{\del r }{\del X^\alpha} \omega_5\,.
\eea
Matching this with (\ref{fieldstr}) requires
\be
V^\alpha \frac{\del r }{\del X^\alpha} = (C' C^5)^{-1}\,,
\ee
which, using the ansatz $V^{\alpha}=\epsilon(r) X^{\alpha}$, is
solved by\footnote{
We used the relation $C^2(r)=X^\alpha X_\alpha$, from which
$\frac{\del r }{\del X^\alpha}=\frac{\del r }{\del C^2(r)}\frac{\del C^2 }{\del X^\alpha}=(CC')^{-1}X^\alpha$.}
\be
V^\alpha = C^{-6} X^\alpha = (1+r^4)^{-\frac{3}{2}}X^\alpha\,.
\ee

We notice that $\epsilon(r \to 0)=\alpha(r \to 0) \to 1$, so that in the near-horizon
 approximation we have $V^{M}=X^{M}$
 %
 %
as was already found in \cite{conffads}.

The general form of the vector field in terms of the 12-dimensional coordinates is
\be
V^{\mu'}=\frac{2X^{\mu'}}{2(X^{\alpha}X_{\alpha})^{3}-X^{M}X_{M}}, \,
V^{\alpha}=\frac{X^{\alpha}}{(X^{\beta}X_{\beta})^{3}}\,.
\ee

\section{Discussion}
The aim of this talk has been to report on a global description \cite{noi}
of non-dilatonic branes by isometrically embedding them in flat
space with two extra dimensions
and two times, thus extending the ideas of \cite{conffads}.
We have gained a rather clear
global picture of the geometry, giving insight in the structure around
coordinate singularities and in the symmetries. In particular, the differences
between $p$-branes
with $p$ even and $p$ odd,  previously pointed out in \cite{GHT},
are clearly apparent.
Like the familiar embedding of anti-de Sitter spacetime as a quadric,
our embeddings are periodic in time.  This is consistent with some suggestions in
\cite{GaryWrap}, but one may of course always pass to the covering space.

In the context of supergravity and string theory, $p$-branes are
coupled to $(p+2)$-form field strengths. An embedding of
the brane thus has to include, besides the embedding of the geometry,
a prescription for the forms in the higher dimensional space. This is
obtained by defining constant $(p+3)$-forms in $D+2$ dimensions, and
contracting them using a vector $V$. The form of $V$ is determined by
matching the projection on the surface with the known forms for the branes
field strengths.

Unfortunately, the geometric significance of the vector field
$V$ remains unclear.
In the case of the M2-brane it is not even
unique, since the $V^\alpha $ components are arbitrary.
A co-dimension 2 surface has a 2-dimensional normal
plane.  In the D3 and M5 cases, the vector $V$ does not lie in this 2-plane, except
in the near-horizon limit. Specifically, the normal 2-plane
is spanned by $\partial _\mu \phi _1$ and $\partial _\mu \phi _2$.
One may check that $V$ is not a linear combination of
$\partial _\mu \phi _1$ and $\partial _\mu \phi _2$.
The bosonic action for probe branes in
the embedded background (\ref{action}) is completely determined after
the construction of $V$.
It could be interesting to investigate if the vector $V$ can have some role in the context of F-theory \cite{vafa}.

Finally it is possible that the methods developed in this paper may
be applicable to scenarios in which one regards the universe as a
brane embedded in a higher-dimensional spacetime.

\section*{Acknowledgments}
\noindent
This work is supported by the European Commission TMR programme
ERBFMRX - CT96 - 0045. C.H is funded by FCT (Portugal) through grant no. PRAXIS XXI/BD/13384/97.


\end{document}